\documentclass[reprint,superscriptaddress,aps,prb,twocolumn,floatfix]{revtex4-2}
\usepackage[utf8]{inputenc}
\usepackage{textcomp} 
\usepackage{setspace}
\usepackage{amsmath}
\usepackage{breqn}
\usepackage{graphicx}
\usepackage{verbatim}
\usepackage{float}

\usepackage{amsfonts}
\usepackage{amssymb}
\usepackage{xcolor}
\usepackage{soul}
\usepackage{tabularx}
\setcitestyle{super}
\usepackage[colorlinks,linkcolor=blue,anchorcolor=blue,citecolor=blue,urlcolor=black]{hyperref}
\DeclareGraphicsExtensions{.pdf,.eps,.png,.jpg,.mps}
\usepackage{booktabs}
\usepackage{makecell}
\usepackage{threeparttable}

\begin{document}

\title{Sub-milliwatt, widely-tunable coherent microcomb generation\\ with feedback-free operation}
\author{ Haowen Shu$^{1,\dagger}$, Lin Chang$^{2,\dagger}$, Chenghao Lao$^{3,\dagger}$, Bitao Shen$^{1,\dagger}$, Weiqiang Xie$^{2}$, Xuguang Zhang$^{1}$, Ming Jin$^{1}$, Yuansheng Tao$^{1}$, Ruixuan Chen$^{1}$, Zihan Tao$^{1}$, Shaohua Yu$^{1,4}$, Qi-Fan Yang$^{3,5}$, Xingjun Wang$^{1,4,5,*}$ and John E. Bowers$^{2,*}$ \\
\vspace{3pt}
$^1$State Key Laboratory of Advanced Optical Communications System and Networks, Department of Electronics, School of Electronics Engineering and Computer Science, Peking University, Beijing, 100871, China.\\
$^2$Department of Electrical and Computer Engineering, University of California, Santa Barbara, CA 93106, USA.\\
$^3$State Key Laboratory for Artificial Microstructure and Mesoscopic Physics, School of Physics, Peking University, Beijing 100871, China.\\
$^4$Peng Cheng Laboratory, Shenzhen 518055, China.\\
$^5$Frontiers Science Center for Nano-optoelectronics, Peking University, Beijing 100871, China.\\
$^\dagger$These authors contributed equally to this work \\
\vspace{3pt}
Corresponding authors: $^*$xjwang@pku.edu.cn, $^*$bowers@ece.ucsb.edu.}



\date{\today}

\maketitle
\noindent
\textbf{Abstract} \\
\textbf{Microcombs are revolutionizing optoelectronics by providing parallelized, mutually coherent wavelength channels for time-frequency metrology and information processing. To implement this essential function in integrated photonic systems, it is desirable to drive microcombs directly with an on-chip laser in a simple and flexible way. However, two major difficulties are preventing this goal: 1) generating mode-locked comb states usually requires a significant amount of pump power and 2) the requirement to align laser and resonator frequency significantly complicates operation and limits the tunability of the comb lines. Here, we address these problems by using microresonators on an AlGaAs on-insulator platform to generate dark-pulse microcombs. This highly nonlinear platform dramatically relaxes fabrication requirements and leads to a record-low pump power of less than 1 mW for coherent comb generation. Dark-pulse microcombs facilitated by thermally-controlled avoided mode-crossings are accessed by direct DFB laser pumping. Without any feedback or control circuitries, the comb shows good coherence and stability. This approach also leads to an unprecedented wide chirping range of all the comb lines. Our work provides a route to realize power-efficient, simple and reconfigurable microcombs that can be seamlessly integrated with a wide range of photonic systems.}

\vspace{3pt}
\noindent
\textbf{Introduction} \\
Microcombs have witnessed great success in the last decade  \cite{stern2018battery,raja2019electrically,shen2020integrated,jin2021hertz,xiang2021laser,Kippenberg2018}. Due to the indispensable advantages of generating coherent, equidistant optical frequency lines in a scalable way \cite{cundiff2003colloquium,diddams2020optical}, they provide tantalizing prospects in a wide range of applications, such as communications, sensing, computations, spectroscopy \cite{suh2016microresonator,yang2019vernier}, optical clocks \cite{newman2019architecture} and frequency synthesis \cite{spencer2018optical}. To facilitate their use in these systems, a major goal in the development of microcombs throughout their whole history has been to generate coherent comb states efficiently, simply and with wide tuning range. Although remarkable advances have been made towards this goal in both fabrication \cite{spencer2014integrated,ji2017ultra,yang2018bridging,liu2020photonic,jin2021hertz,liu2021720,Ye1,Ye2} and operating principle \cite{Generation1, Generation3, shen2020integrated}, so far there is still no approach that can firmly unite all these properties. This obstacle prevented the spread and further commercialization of microcomb technologies, but has now been solved.

One essential problem is the high pump power requirement. In particular, for miniaturization and ultimately high-volume production of integrated systems, microcombs need to be directly pumped by an on-chip laser, whose pump power is usually in the tens of milliwatts range. In order to generate combs efficiently in this power regime, tremendous effort have been spent on the improvement of microresonator quality factors ($Q$) over the last decade by fine-tuning fabrication recipes. So far, the Si$_3$N$_4$ platform is leading in this regard, thanks to its extraordinary $Q$, now up to tens of millions for thick Si$_3$N$_4$ microresonators \cite{Fab1} and hundreds of millions for thin ones \cite{jin2021hertz}. The required pump power for coherent comb states can be reduced to below 20 mW, which can be reached by an on-chip distributed feedback laser (DFB) or Fabry-Perot (FP) laser diode \cite{stern2018battery,raja2019electrically,shen2020integrated,jin2021hertz,xiang2021laser}. However, such advances in resonator $Q$ rely on many extra fabrication procedures, including chemical-mechanical polishing and high temperature annealing. These procedures add cost to the fabrication. 

\begin{figure*}[ht]
\centering
\includegraphics[width = 18cm]{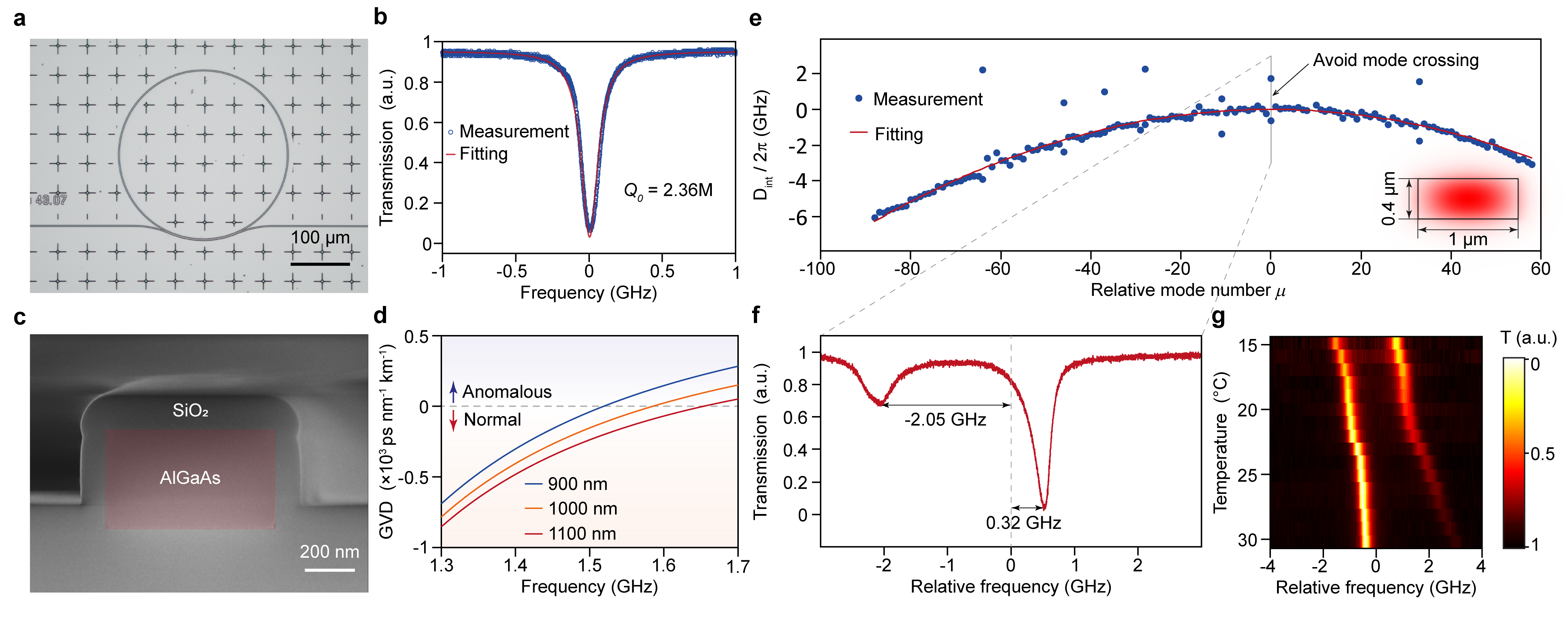}
\caption{\textbf{Resonator characterization. a,} Top-view photograph of the AlGaAsOI microresonator. The radius of the microresonator is 144 \textmu m, corresponding to 90 GHz free spectral range (FSR). Scale bar: 100 \textmu m \textbf{b,} Normalized transmission spectrum of a typical resonance at 1552 nm. Lorentzian fitting reveals intrinsic $Q$ factors about 2.36 million. \textbf{c,} Scanning-electron-microscopy (SEM) image showing the cross section of the microresonator. The AlGaAs core is highlighted in red, and silica forms the substrate and cladding. Scale bar: 200 nm {\bf d,} Calculated dispersion of the TE0 mode in AlGaAsOI microresonators with respect to the width of the core. The thickness of the core is set as 400 nm. The inset shows the corresponding TE0 mode profile. {\bf e,} Measured TE0 mode family dispersion. $D_{\rm int}=\omega_\mu-\omega_o-D_1\mu=D_2\mu^2/2+\mathcal{O}(\mu^3)$, where $\omega_\mu$ is the resonant frequency of the $\mu_{\rm th}$ mode. The index of the mode that is pumped  is set to 0, and $D_1$ is the free-spectral-range at pump wavelength. Parabolic fitting (red) shows $D_2/2\pi=-1.63$ MHz. {\bf f,} Zoom-in spectrum of the doublet resonances at $\mu=0$ indicated in panel {\bf e}. The frequency is plotted relative to the fitting curve. {\bf g,} Pseudo-color plot of the normalized transmission spectra at $\mu=0$ as a function of temperature.}
\label{fig1}
\end{figure*}

Another key challenge comes from the operation side: Nowadays, the mainstream work of microcombs leverages dissipative Kerr solitons (DKS, so-called bright soliton) for coherent comb generation. Since these states exist at the red detuning side of the resonance of the cavity \cite{herr2014temporal}, they are thermally unstable and therefore require special tricks to align the pump laser and the resonances in soliton formation.  The most commonly-used strategies are currently active capture techniques and delicate tuning \cite{Generation1,Stabilize2}. These approaches require benchtop laser sources and complex control protocols, which are not suitable for integrated photonic systems. Recent demonstrations of direct pumping based on self-injection locking have enabled turnkey soliton operation by directly coupling an on-chip laser into a microresonator, which gets rid of all the tuning procedures and greatly simplifies the microcomb generation scheme \cite{raja2019electrically,shen2020integrated,xiang2021laser}. However, such coupling relies on either hybrid or heterogeneous integration, which demands significant effort in process development and hasn’t yet been integrated into existing photonic infrastructures. Importantly, all of the approaches discussed above need feedback, either optical or electrical, to lock the laser frequency within the access window of a coherent state that is usually quite narrow (usually within 1 GHz for bright soliton and MHz level for injection locking). Hence, they all suffer from limited modulation range of comb lines. These drawbacks affect the performance in many essential applications such as Lidar \cite{riemensberger2020massively} and spectroscopy \cite{Spec1}.

In this work, we address this long-standing problem by generating dark-pulse microcombs on a highly nonlinear photonic platform, Aluminum Gallium Arsenide-on-Insulator (AlGaAsOI). Thanks to the giant nonlinearity of the material, AlGaAsOI microresonators support coherent comb generation with an unprecedented low pump level of 930 \textmu W with moderate $Q$ value, which significantly relaxes the fabrication requirements. By controlling the avoided-mode-crossings (AMXs) in AlGaAsOI microresonators with normal dispersion, dark-pulse microcombs can be readily generated without electronic feedback control. The strong thermal nonlinearity of AlGaAs also facilitates stable operation of microcombs by creating a wide window (up to 11 GHz) for the desired coherent state. This appealing feature enables direct pumping of AlGaAsOI dark-pulse by a commercial DFB laser diode, as well as delivering a record high chirping range for all comb lines with good uniformity. The phase and intensity noise of a free-running comb across the entire C-band are verified, which shows short-term linewidth of several kHz and the average RIN down to -135 dB Hz$^{-1}$, and with more than 7 hours stable operation. This approach provides an appealing microcomb solution for a wide range of system-level optoelectronic applications and holds great potential for fully integrated photonic systems in the future.

\begin{figure*}[ht]
\centering
\includegraphics[width = 18cm]{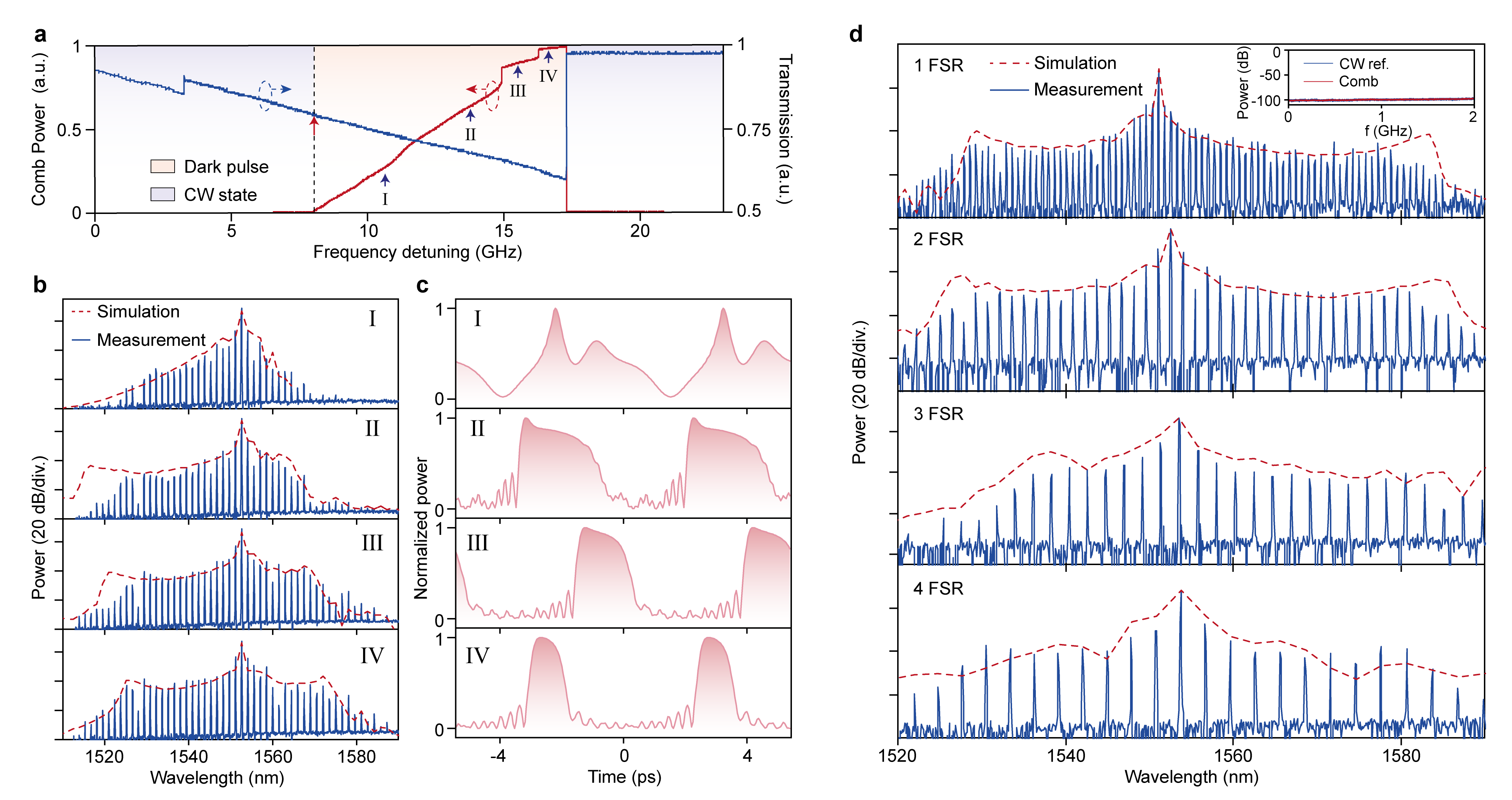}
\caption{\textbf{Dynamics of mode-locked microcombs. a,} Normalized transmitted total power (blue) and comb power (red). The laser is scanned from the blue side to the red side of the mode. Blue and red shadings indicate continuous-wave (CW) state and microcomb state, respectively. {\bf b,} Typical optical spectra of microcombs at different stages as indicated in {\bf a}. Red dashed lines denote the simulated spectral envelope. {\bf c,} Simulated intracavity waveforms corresponding to the spectra in {\bf b}. {\bf d,} Optical spectra of microcombs with spacing from 1-4 FSRs (top to bottom). Inset: intensity noise of the 1-FSR microcomb (resolution bandwidth: 100 kHz). The noise floor of the measurement system is also plotted for comparison.}
\label{fig2}
\end{figure*}

\vspace{3pt}
\noindent
\textbf{Results} \\
\textbf{Dark-pulse in an AlGaAsOI microresonator} \\
\noindent The AlGaAsOI microresonator used in this work is shown in Fig. \ref{fig1}a. The epitaxial AlGaAs is transferred onto a silicon substrate with a 3-\textmu m thick thermal silica layer via a wafer bonding process. With optimal fabrication process \cite{chang2020ultra,xie2020ultrahigh}, intrinsic $Q$ factors over 2 million are achieved (see Fig. \ref{fig1}b), corresponding to waveguide propagation loss of 0.3 dB/cm. Figure \ref{fig1}c shows the cross section of the waveguide, where the thickness of the AlGaAs core is 400 nm. With this geometry, the TE0 mode in a 1 \textmu m width waveguide is anticipated to feature normal dispersion around 1550 nm based on numerical simulation (see Fig. \ref{fig1}d). Experimental measurement of the TE0 mode family dispersion from 1510 nm to 1630 nm is plotted in Fig. \ref{fig1}e, which explicitly shows normal dispersion. Hybridization of TE0 modes and neighboring modes is observed at certain wavelengths, and the resulting AMX expels the resonances from the parabolic fitting curve. The hybrid mode, which is red shifted from its original position, induces local anomalous dispersion and initiates parametric oscillation when it is sufficiently pumped \cite{xue2015mode,lobanov2015frequency,kim2019turn,helgason2021dissipative}. The dispersion spectrum measured at 23.1 $^{\circ}$C) shows that a hybrid mode is red-shifted by 0.32 GHz due to AMX (see Fig. \ref{fig1}f). 

\begin{figure*}[ht]
\centering
\includegraphics[width = 18cm]{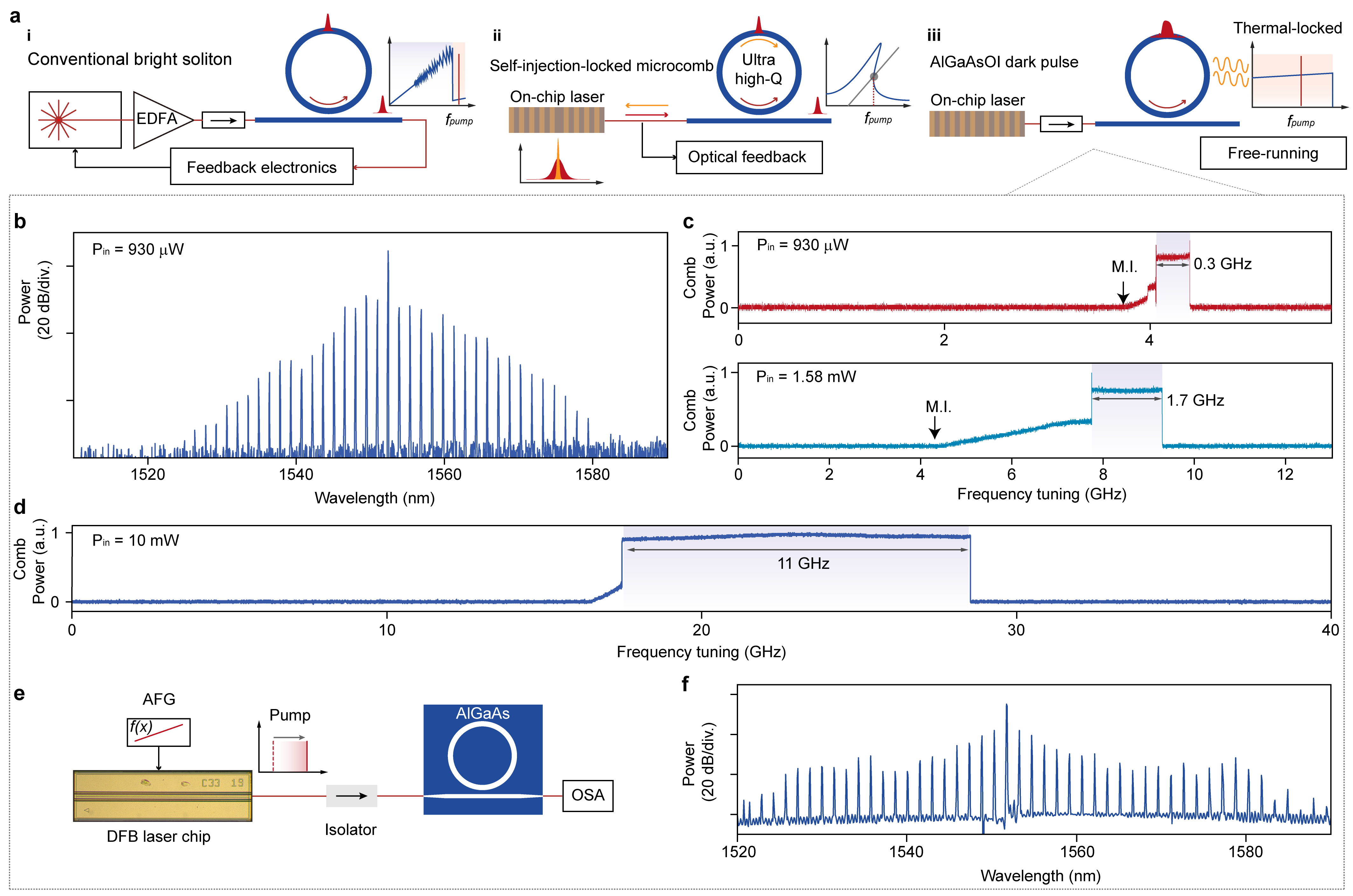}
\caption{\textbf{Power-efficient mode-locked Kerr comb generation. a,} Schematics of different microcomb generation schemes. \textbf{b,} Comb spectrum at the flat step (shading area in upper panel (\textbf{c})) under the on-chip pump power of 930 \textmu W. \textbf{c,} Measured comb power with respect to the frequency tuning. As the increasing of the pump power, the comb existence area could be extended from 0.3 GHz (upper panel) to 1.7 GHz. \textbf{d,} Power transmission of a microcomb with 11 GHz flat step under the same power level (10 mW) as an off-the-shelf distributed feedback (DFB) laser can offer. \textbf{e,} Experimental set-up of an on-chip semiconductor laser pumped scheme and the \textbf{f,} comb spectrum. All the spectra are of 2-FSR ($\sim$180 GHz) frequency spacing. AFG: arbitrary function generator; OSA: optical spectrum analyzer.
}
\label{fig3}
\end{figure*}

The temperature of the microresonator is controlled using a thermoelectric cooler underneath the chip. The resulting frequency shift of a certain mode ($\Delta\omega$) as a function of the temperature change ($\Delta T$) yields
\begin{equation}
    \Delta\omega=-\frac{n_T}{n_{\rm eff}}\omega\Delta T,
\end{equation}

where $n_{\rm eff}$ and $\omega$ denote the effective index and resonant frequency of the mode, respectively, and  $n_T$ is the thermo-optic coefficient of the material. Due to the distinct mode confinement of neighboring mode, their frequency difference could change accordingly as a function of temperature. Figure \ref{fig1}g shows the normalized transmission of the hybrid modes near $\mu=0$ measured at different temperatures. It is noted that the relative frequency of the left branch is shifted towards the red side as the temperature decreases. Moreover, Since the $n_T$ of AlGaAs (2.3 $\times$ 10$^{-4}$ K$^{-1}$) is an order of magnitude larger than those of silica and Si$_3$N$_4$, the on-chip thermo-tuning efficiency in AlGaAsOI can be much higher ($\sim$5 GHz/$^{\circ}$C) than those platform , showing great efficiency for AMX operation \cite{TermoTune}.

The microresonator with is first pumped by an external-cavity diode laser for microcomb generation. The waveguide dimension is 1 \textmu m $\times$ 400 nm. With 5 mW optical power launched into the bus waveguide, frequency comb formation is observed when the laser is manually tuned into a mode, as shown in Fig. \ref{fig2}a. The transmission of comb power features 4 steps associated with different states of microcombs \cite{Elhan2021}. Unlike the case of bright solitons in microresonators with anomalous dispersion \cite{herr2014temporal,yi2015soliton,brasch2016photonic,stone2018thermal,moille2019kerr,moille2020dissipative}, abrupt change of intracavity power is not as significant here. Fig. \ref{fig2}b compares the optical spectra of the microcombs at different stages, whose spans are broadened as the detuning between the laser and mode increases. The spacing between adjacent comb lines is two FSRs of the microresonator. At stage III and IV, a pair of flat wings are formed on both sides of the pump, which is a signature feature of dark-pulses in microresonators \cite{xue2015mode}. The asymmetry of the spectra results from AMX \cite{Elhan2021}, which is further confirmed using numerical simulations based on coupled Lugiato-Lefever equations (see Methods and Supplementary Note II). The simulated temporal waveforms of the microcombs are plotted in Fig. \ref{fig2}c. It is noted that the intracavity waveforms could evolve into bright pulses when the detuning is sufficiently large, which agrees well with previous reports \cite{helgason2021dissipative,Elhan2021,jin2021hertz}.

Microcombs with perfectly multiple-FSR spacing exhibit enhanced power per comb line \cite{cole2017soliton,karpov2019dynamics,lu2021synthesized}. By adjusting the temperature and pump wavelength, microcombs with varying FSRs are generated, whose optical spectra are shown in Fig. \ref{fig2}d. The microcomb with a single FSR spacing is photodetected and sent into an electrical spectrum analyzer (ESA). The measured intensity noise of the entire microcomb coincides with the measurement floor, which confirms the mode-locking nature of the microcomb. The remarkable operability of multi-FSR dark-pulses may enjoy great potential at all-to-all entanglement quantum source \cite{SCforQuantum}, tunable radio-frequency filter \cite{SCforRFFilter} and THz carrier generation \cite{SCformmWave}.

\begin{figure*}[ht]
\centering
\includegraphics[width = 14cm]{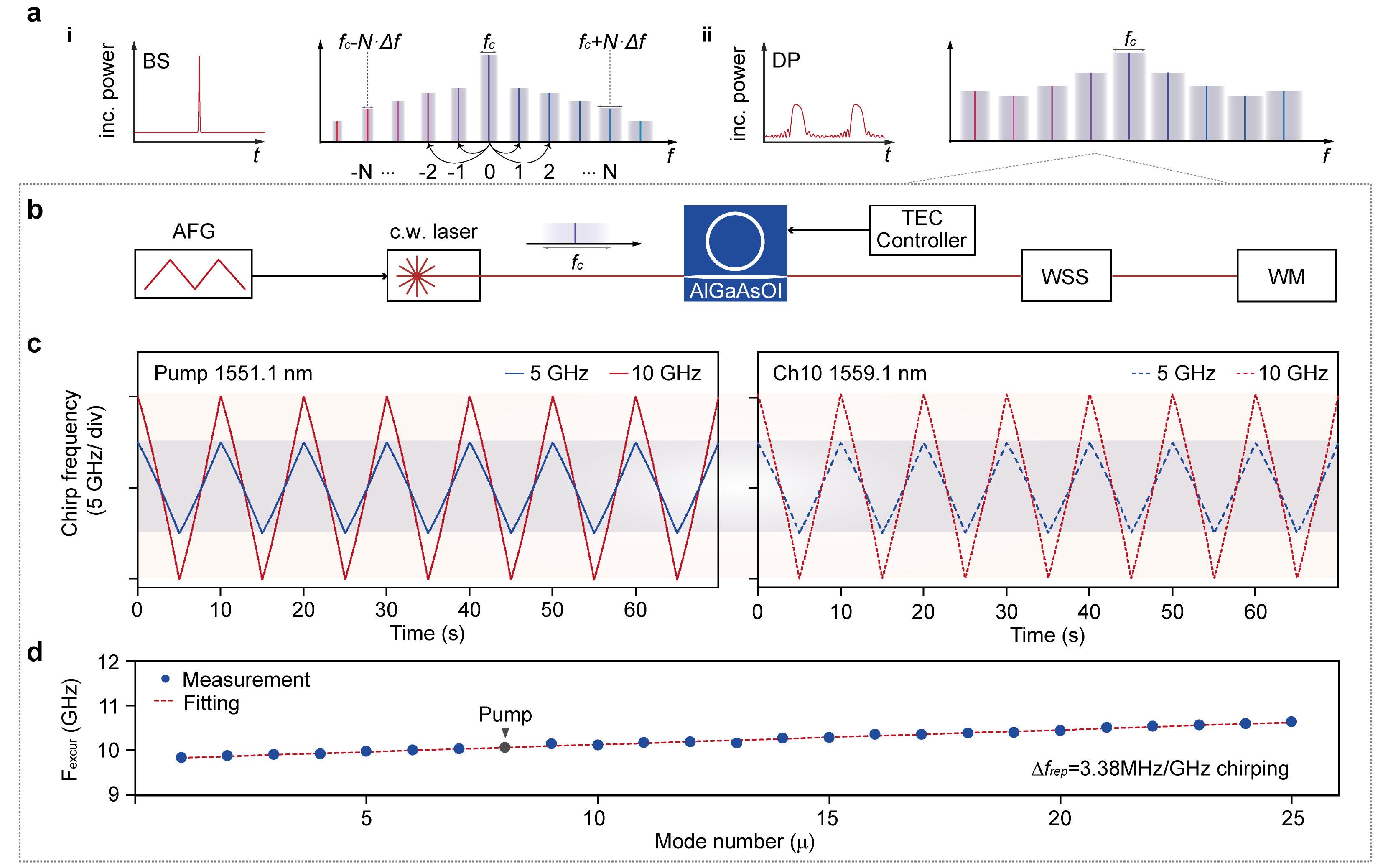}
\caption{\textbf{Widely frequency chirped microcomb. a,} Both (i) bright soliton and (ii) dark-pulse could act as a parallel frequency chirping source, in which the frequency modulation of the pump laser is transduced to each comb line. The stimulated Raman effects and higher order dispersion would result in wavelength-dependent chirping copies, which could be mitigated in the efficient dark-pulse scheme with relatively low intracavity power. \textbf{b,} Experimental setup of the parallel chirping source. WSS: wavelength selective switch; WM, wavelength meter. \textbf{c,} Measured time-frequency maps of the pump line (left panel) and the channel-10 sideband (right panel), with the pump laser chirping at 5 GHz (red) and 10 GHz (blue), respectively. \textbf{d,} Frequency excursion of each channel at 10 GHz frequency chirping.
}
\label{fig4}
\end{figure*}

\vspace{3pt}
\noindent\textbf{Direct-pumping of coherent microcombs}\\
\noindent A critical step towards an integrated laser-microcomb source is the realization of a mode-locked microcomb pumped by an on-chip laser, which requires both high efficiency for comb generation and simple operation strategies. In addition to the recently demonstrated laser-soliton microcomb based on injection-locking, the dark-pulse comb in AlGaAsOI microresonator gives another solution \cite{pu2016efficient,chang2020ultra}. In our experiment, microcomb generation is still feasible when the pump power is reduced to the sub-milliwatt-level (930 \textmu W), featuring nearly 40 spectral lines spanning over the entire C band (see Fig. \ref{fig3}b). The operation power is much smaller than previously reported coherent comb results. At such low power level, a single DFB can drive potentially tens of microcombs, which is invaluable in applications such as communications or dual comb spectroscopy. 

\begin{figure*}[ht]
\centering
\includegraphics[width = 18cm]{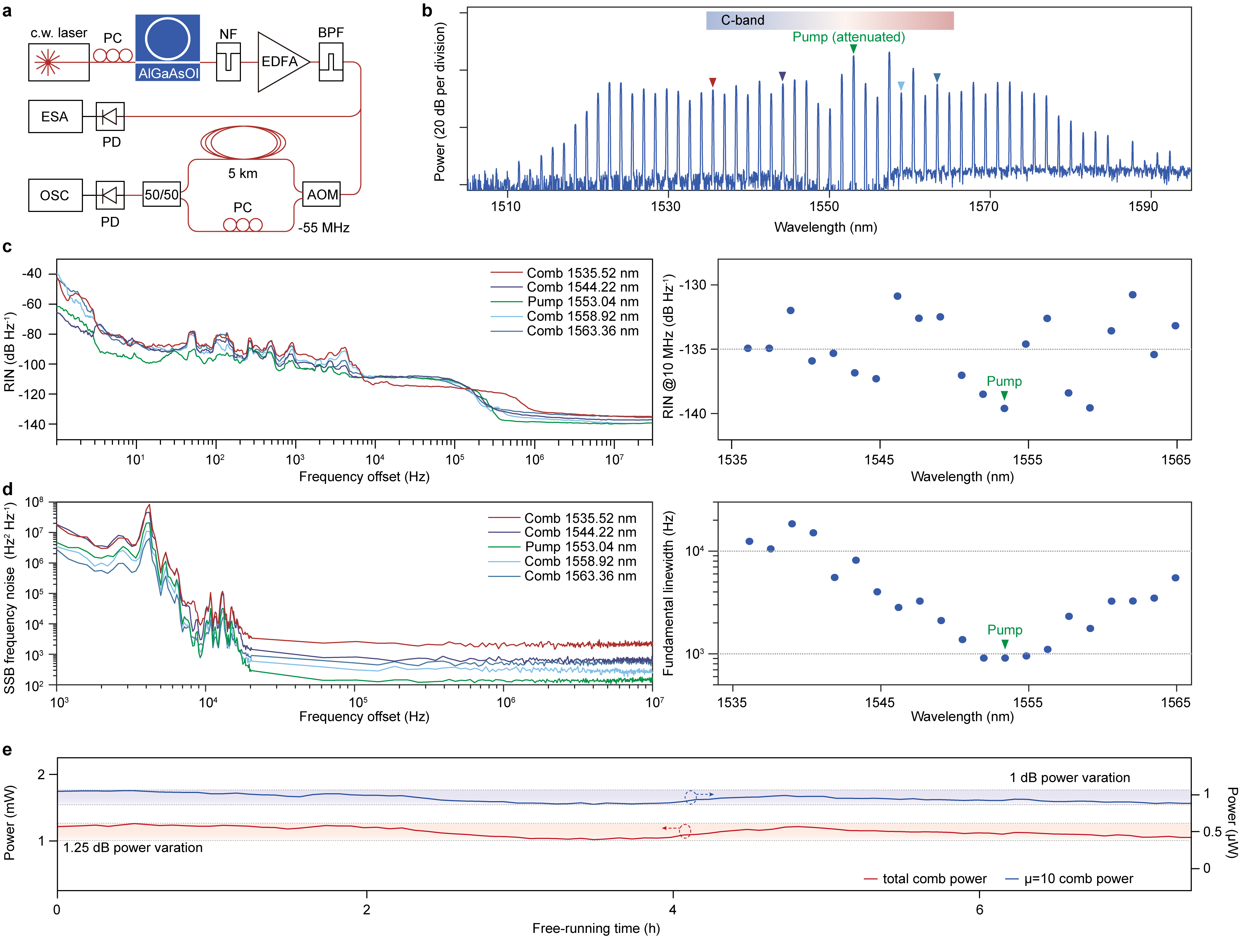}
\caption{\textbf{Coherence of mode-locked frequency comb.} \textbf{a,} Experimental setup. PC: polarization controller; NF: notch filter: EDFA, erbium-doped fiber amplifier; BPF: band-pass filter; AOM: acousto-optic modulator; PD: photodetector; ESA: electric spectrum analyzer; OSC: oscilloscope. \textbf{b,} Optical spectra of dark-pulse comb after notch filter. The range of telecommunication C-band is also indicated. \textbf{c,} Left panel: measured relative intensity noise (RIN) of comb teeth indicated in {\textbf b}. Right panel: RIN at 10 MHz offset frequencies of all comb teeth within C-band. {\textbf d,} Left panel: measured single-sideband (SSB) frequency noise of comb teeth indicated in {\textbf b}. Right panel: fundamental linewidth of all comb teeth within C-band. \textbf{e,} Long term stability of the microcomb. The optical spectra of the microcomb are continuously recorded for over 7 hours, and the total comb power (red) and power of the 10th comb line (blue) are plotted.}
\label{fig5}
\end{figure*}

In general, a clean “step” of the comb power transmission usually indicates an accessing range of mode-locked comb with small power deviation.  For dark-pulse state, these steps are significantly longer than those of bright solitons due to the thermal stability at blue detuning. And this nature can be further enhanced on AlGaAsOI due to the strong Kerr and thermo-optic effects, which significantly extend the thermal triangle of the response of the cavity. The transmission comb power under different pump powers is recorded in Fig. \ref{fig3}c and d, where the “step length” gets extended as the increasing of pump power. With only 1.58 mW on chip power, the step size we got has already been around that in bright soliton generation under several hundreds of mW pump power \cite{riemensberger2020massively}. At 10 mW, it exhibits an 11 GHz soliton plateau. Such a long existence range ensures resilience to frequency vibration of the pump laser, particularly for the on-chip lasers whose stability are much worse than external-cavity based narrow-linewidth lasers. As a result, this scheme gets rid the need of any kind of feedback to lock the laser onto the cavity (see Fig. \ref{fig3}a). For proof-of-concept, we choose a commercial distributed feedback (DFB) laser to drive the microresonator, which is coupled to the bus waveguide via a lensed fiber, as depicted in Fig. \ref{fig3}e. An optical isolator is inserted between the laser and the microresonator to reject backreflected signals. The optical spectrum of the 2 FSR-spaced microcomb is plotted in Fig. \ref{fig3}f. This configuration, compared to injection locking, doesn’t require a specialized chip-to-chip packaging or heterogeneous laser integration, but can be easily achieved via current standard fiber-to-chip packaging process in integrated photonic industry. Therefore, it can be seamlessly implemented in a wide range of existing photonic systems.

\vspace{3pt}
\noindent\textbf{Widely-tunable massively parallel source}\\
\noindent The large step and feedback-free nature offers great frequency chirping capability (see Fig. \ref{fig4}) to this microcomb generation scheme. This is of great importance in many essential applications such as Lidar, as the chirping range of the frequency combs is inversely proportional to the fundamental ranging resolution in frequency-modulated continuous-wave (FMCW) scheme \cite{riemensberger2020massively,FMCW1,FMCW2}. Fig. \ref{fig4}b shows the setup to characterize frequency chirping, with details described in Methods. As a proof-of-concept demonstration, a triangular frequency modulation scan (5 GHz and 10 GHz) is directly applied to the pump laser acting as a FMCW source (see left panel of Fig. \ref{fig4}c).  When the comb operates within the dark-pulse step, the modulated frequency can be transduced to all comb teeth, resulting in a 30-channel parallel FMCW source with $>$10 GHz frequency excursion bandwidth (see right panel of Fig. \ref{fig4}c and Supplementary Note III). Such frequency chirping range corresponds to a sub-centimeter-level ranging resolution, which is one order of magnitude better than the previous microcomb-based parallel ranging result \cite{riemensberger2020massively}.

The repetition frequency shift induced by the stimulated Raman effect \cite{Raman1} and higher order dispersion \cite{DW1,DW2} is a problem in bright soliton comb during the frequency chirping. It causes inconsistencies of the chirping range over different comb lines (see Fig. \ref{fig4}a). This can be greatly alleviated in our approach due to the low pump power and large duty cycle of dark-pulse circulating in the cavity that cause lower peak power. The channel-dependent frequency excursion of each mode is depicted in Fig. \ref{fig4}e, showing a repetition rate mismatch of only 3.38 MHz per GHz of laser tuning, about 4 times smaller than that in the previous result. Such consistency results in more uniform measurement errors among channels, improving the overall ranging resolution.

\begin{table*}[ht]
	\centering
	\caption{Comparison of coherent microcomb generated with various integrated nonlinear platforms}
	\begin{tabular}{p{1.5cm}p{2.6cm}<{\centering}p{1.7cm}<{\centering}p{1cm}<{\centering}p{2cm}<{\centering}p{2cm}<{\centering}p{1.5cm}<{\centering}p{4.5cm}}
	
		\toprule  
		Platform & Comb type & \makecell[c]{$Q$-factor \\ (Million)} & \makecell[c]{FSR \\ (GHz)}  & \makecell[c] {On-chip \\ power (mW)} &
		\makecell[c] {Soliton step \\ (GHz)} &
		\makecell[c] {on-chip \\ laser} &
		Accessing method \\ 
		\midrule 

Silica\cite{yang2018bridging} & bright soliton & $>$200 & 15 & 25 & - & N & Power kicking  \\
Si$_3$N$_4$\cite{CP1} & bright soliton & 15 & 99 & 6.2 & - & N & Frequency scanning  \\
Si$_3$N$_4$\cite{li2017stably} & bright soliton & 1.4* & 230 & 200 & $\sim$1.5 & N & Frequency scanning  \\
Si$_3$N$_4$\cite{shen2020integrated} & bright soliton & 16 & 15 & 30 & N/A & Y & Injection locking  \\
GaN\cite{CP4} & bright soliton & 1.8 & 324 & 136 & - & N & \makecell[l] {Manual frequency tuning\\ (Auxiliary laser)}  \\
LiNbO$_3$\cite{CP5} & bright soliton & $\sim$1.1* & $\sim$200 & $\sim$90 & $\sim$0.5 & N & Bi-directional scaning  \\
LiNbO$_3$\cite{CP6} & bright soliton & 2.2* & 199.7 & 33 & $\sim$0.5 & N & Bi-directional scaning  \\
SiC\cite{SCforQuantum} & bright soliton & 5.6 & 350 & 2.3 & $\sim$0.014 & N & Frequency tuning in cryostat \\
Hydex\cite{CP8} & soliton crystal & 1.5 & 48.9 & $\sim$1100 & - & N & Manual frequency tuning \\
AlN\cite{CP9} & bright soliton & 1.4 & 374 & $\sim$335 & 10.4 & N & \makecell[l] {Manual frequency tuning\\ (Auxiliary resonance)}  \\
AlN\cite{CP10} & bright soliton & 1.6 & 433 & $\sim$390 & $\sim$10 & N & Frequency scanning  \\
Ta$_2$O$_5$\cite{CP11} & bright soliton & 0.4* & 1000 & 36 & $\sim$3 & N & Frequency scanning  \\
Si$_3$N$_4$\cite{helgason2021dissipative} & dark-pulse & 5.7 & 104.8 & 2.5 & $\sim$0.25 & N & Frequency scanning  \\
Si$_3$N$_4$\cite{xue2015mode} & dark-pulse & 0.77* & 231.3 & $\sim$850 & $\sim$37.5 & N & Thermally tuned resonance  \\
Si$_3$N$_4$\cite{jin2021hertz} & dark-pulse & 260 & 5 & $\sim$20 & N/A  & Y & Injection locking  \\
Si$_3$N$_4$\cite{2021Platicon} & dark-pulse & $>$10 & 26.2 & 5 & N/A  & Y & Injection locking  \\
\makecell[l]{AlGaAsOI\\ (this work)} & dark-pulse & 2.36 & 180 & 0.93 & 11 & Y & Manual frequency tuning  \\
		\bottomrule  
	\end{tabular}
   \begin{tablenotes}
     \item* represents the loaded $Q$.
   \end{tablenotes}
\label{tbl1}
\end{table*}

\vspace{3pt}
\noindent\textbf{Coherence and stability of free-running microcomb}\\
\noindent The stability and coherence of microcombs are essential in almost all applications. Previously, all these characterizations of microcombs are under certain locking schemes. Here, for the first time, we measure the RIN noise, phase noise and long-term stability for a free-running dark-pulse microcomb. The experimental setup of noise evaluation is illustrated in Fig. \ref{fig5}a, more details can be found in Methods. The measured 2-FSR microcomb whose optical spectrum shown in Fig. \ref{fig5}b spans over 80 nm and features a relatively flat envelope within the telecommunication C-band. RIN measurements of the comb teeth are plotted in Fig. \ref{fig5}c. At low offset frequencies the noise is primarily induced by thermal and mechanical instability and by the optical amplifier. For offset frequencies greater than 1 MHz, the RIN maintains constant; the flatline value indicates the intrinsic noise characteristics of the microcomb and is of interest to applications such as data links. The intrinsic RIN of comb lines within the C-band are measured and are all below -130 dB Hz$^{-1}$, with a few approaching -140 dB Hz$^{-1}$. 

The frequency noise of the comb lines is shown in Fig. \ref{fig5}d, which converges to white noise above 20 kHz offset frequencies. The fundamental linewidth of the pump is about 1 kHz. Decoherence of comb lines is observed as their linewidth increases quadratically with their spectral separation from the pump, which can be attributed to the instability of the repetition rate due to the free-running operation mode. Nonetheless, the derived fundamental linewidth of the comb lines within C-band are still at the level of a few kilohertz, which is suitable for coherent communications, high-precision sensing and metrology.

This free-running microcomb shows stable operation over 7 hours (see Fig. \ref{fig5}). By actively recording the optical spectra of the microcomb every 5 minutes, the power drift of the microcomb is obtained, which only varies within 1.25 dB for the total comb power, and 1 dB for certain comb lines. The instability of the coupling setup is suspected to cause this drift, and can be significantly suppressed once the device is packaged.

\vspace{3pt}
\noindent
\textbf{Discussion} \\
A comparison of different integrated nonlinear platforms for coherent microcomb generation is shown in Table \ref{tbl1}. In this work, the pump power used is the lowest number required for coherent microcomb generation. Combined with the feedback-free operation and wide tunability, this approach, to our best knowledge, is the only one that attain all these desirable features at the same time.

Although the laser is free-running, its stability of phase and amplitude of the comb lines is sufficient to incorporate complex signal processing and transmission schemes, e.g., 4 pulse amplitude modulation (PAM4) and 16 quadrature amplitude modulation (16QAM) \cite{liu2019high,marin2017microresonator}. The maximum power conversion efficiency we achieved is around 15 $\%$, which is more than an order of magnitude higher than that of typical bright soliton microcombs. This efficiency can be further increased by improving coupling or introducing coupled-microresonator geometry for dynamic adjustment of AMXs \cite{xue2015normal,kim2019turn,helgason2021dissipative}. Previously, at anomalous dispersion regime, only a special class of bright solitons, soliton crystals, exhibit similar robustness and operation simplicity as those of our work here without relying on any feedbacks. These properties are invaluable in practical applications,  including high-volume data transmission \cite{corcoran2020ultra}, optical neural networks\cite{xu202111}, microwave photonics systems\cite{xu2019microcomb}, etc.    

Our results represent a key milestone towards laser-integrated microcomb sources favorable for tremendous system-level applications. The compact and robust microcomb module comprising a DFB laser and the AlGaAsOI microresonator has for the first time enabled the union of microcomb and silicon photonic engines for optoelectronic systems \cite{shu2021Bridging}. Moving forward, with the assistance of an on-chip isolator, chip-to-chip coupling of the laser and microresonator will further improve the integration level of the system \cite{stern2018battery,raja2019electrically,shen2020integrated,jin2021hertz}. It is worth noting that monolithic integration of the AlGaAsOI microresonators with III-V gain sections are feasible given the excellent compatibility of the joint fabrication process, which is considerably simpler than the current heterogeneously integrated laser-soliton microcomb technology. Overall, the realization of ultra-efficient coherent microcombs with wide tunability and operation simplicity in high-$Q$ III-V microresonators provide a promising solution for microcomb generation, and it will accelerate the adoption of frequency comb sources in practical applications such as Lidar, data transmission and optical neural networks \cite{corcoran2020ultra,riemensberger2020massively,feldmann2021parallel}.

\vspace{3pt}
\noindent \textbf{Methods}\\
\begin{footnotesize}
\noindent \textbf{Design and fabrication of the devices. } 
The pulley-type AlGaAsOI resonators were designed to be slightly over coupled and exhibit normal dispersion within C-band. An inverse taper with the waveguide width adiabatically narrowed to 200 nm is used here for efficient chip-to-fibre coupling, the coupling loss is $\sim$2-3 dB/facet. The fabrication of AlGaAs microresonators was based on heterogeneous wafer bonding. Molecular-beam epitaxy (MBE) growth method was employed for the AlGaAs epitaxial wafer. A 248 nm deep-ultraviolet (DUV) stepper was used for the lithography. A photoresist reflow process and an optimized dry etch process were applied in waveguide patterning for waveguide scattering loss reduction. For passivation, the core is fully clad by consecutive deposition of silica via atomic layer deposition (ALD) and plasma enhanced chemical vapor deposition (PECVD). The detailed fabrication process can be found in our previous works at \cite{chang2020ultra,xie2020ultrahigh}.

\vspace{3pt}
\noindent\textbf{Experimental details.} 
The $Q$ factor measurement is performed by scanning an external-cavity-diode laser (Toptica CTL 1550) across a resonance from the blue side to the red side. To avoid the distortion of lineshape caused by thermal nonlinearity, the on-chip power is reduced to tens of microwatts. The dispersion of the resonance is obtained by scanning the laser over a larger wavelength range (from 1520 nm to 1630 nm) and recording the transmission spectrum. The frequency of the laser is referenced to a Mach-Zehnder interferometer, which is calibrated by a fiber comb. 

For RIN noise characterization, the RIN of a low-noise continuous-wave laser is first tested to evaluate the measurement floor, whose power nearly saturates the photodetector (Newport 1811-FC). The measurement floor at high-offset frequency is around -140 dB Hz$^{-1}$, which is primarily limited by the noise-equivalent-power (NEP) of the photodetector. For comb-line measurement, the microcomb is generated with the temperature of the microresonator adjusted to 20.8 $^{\circ}$C.  Then a notch filter is used to attenuate the strong pump and a pair of adjacent comb lines before the comb is sent into an erbium doped fiber amplifier (EDFA). Individual comb lines are then selected using a tunable bandpass filter, and their RIN is characterized by a phase noise analyzer (Rohde Schwarz FSUP 26.5) using the baseband measurement function.

A delayed self-heterodyne setup is used to measure the frequency noise of the comb lines. The setup consists of a 5-km-long fiber delay line. The AOM is driven by a low-noise microwave source at 55 MHz, while the zero-order and first-order signals are sent into the two arms of the interferometer before they are recombined and detected. The signals are recorded using an oscilloscope (Keysight MXR404A), and are Hilbert transformed to extract the instantaneous frequency fluctuations for computation of power spectral densities. To omit the etalon effect, for high offset frequencies ($>$20 kHz), the noise are plotted at certain frequencies given by $f=(2n-1)/2\tau$, with $n$ positive integers and $\tau$ the relative temporal delay between the two arms.

The frequency chirping of the microcomb is performed by directly modulating the pump frequency using an AFG to drive the PZT unit of the external-cavity diode laser. The pump wavelength is around 1551.1 nm, with a microresonator temperature of 16$^{\circ}$C. The pump laser is first manually tuned into the resonance and stopped at the center of the transmission step. Then the triangular frequency modulation of the pump laser is turned on. A wavelength selective switch is employed to filter out each chirped comb line, which is recorded by a high precision wavelength meter (HighFinesse WS6 Series). The period of the symmetric triangular frequency-modulation signal is set to 10 s to satisfy the measurement repetition rate of the wavelength meter, corresponding to the frequency scanning speed of 2 GHz/s.

\vspace{3pt}
\noindent\textbf{Numerical simulation.} 
In order to give a better insight into the self-stimulation of the dark-pulse in our devices, two mode families are considered here. One mode family is the dark-pulse-supporting mode family, referred as primary ($P$) mode family, and the other mode family is referred as auxiliary ($A$) mode family. The avoided-mode-crossing between the two mode families strongly modified the local dispersion, facilitating the self-stimulation of dark-pulses \cite{liu2014investigation,xue2015mode}. The linear coupling between the primary modes and auxiliary modes are introduced into the LLEs:

\begin{equation}
\begin{gathered}
t_{R} \frac{\partial E^{(P)}(t, \tau)}{\partial t}=\left[-\left(\alpha^{(P)}-i t_{R} \delta\right)+i L \frac{\beta_{2}^{(P)}}{2} \frac{\partial^{2}}{\partial \tau^{2}}\right] E^{(P)}
\\+i L \gamma^{(P)}\left|E^{(P)}\right|^{2} E^{(P)}+i L \kappa E^{(A)}+\sqrt{\theta} E_{i n}
\end{gathered}
\end{equation}
\begin{equation}
\begin{gathered}
t_{R} \frac{\partial E^{(A)}(t, \tau)}{\partial t}=\left[-\left(\alpha^{(A)}-i t_{R} \delta-i \Delta\right)+i L \frac{\beta_{2}^{(A)}}{2} \frac{\partial^{2}}{\partial \tau^{2}}\right] E^{(A)}
\\+i L \gamma^{(A)}\left|E^{(A)}\right|^{2} E^{(A)}+i L \kappa E^{(P)}
\end{gathered}
\end{equation}

$E^{(P)}$ and $E^{(A)}$ respectively stand for the intracavity temporal fields in the primary and the auxiliary modes, $\alpha^{(P)}=0.0067$ and $\alpha^{(A)}=0.02$ are the roundtrip cavity loss factor, $\beta_{2}^{(P)}$=139 ps$^{2}$ km$^{-1}$ and $\beta_{2}^{(A)}$=-2421 ps$^{2}$ km$^{-1}$ represent the second-order dispersion coefficients, and $\delta=\omega_{0}^{(P)}-\omega_{p}$ is the detuning, where $\omega_{0}^{(P)}$ is the resonance frequency of the primary mode and $\omega_{p}$ is the frequency of the pump field. $t_{R}$=11.628 ps is the roundtrip time of the primary mode and $L$=2$\pi$×144 $\mu$m is roundtrip length. The pump filed is coupled into the primary mode by $\sqrt{\theta} E_{\text {in }}$, where $\theta$=0.0067 is the waveguide coupling coefficient and $E_{in}$ is the pump field. While the coupling between the pump field and the auxiliary mode is ignored, due to the relatively small coupling rate in the pulley couplers. $\gamma^{(P)}$=340 m$^{-1}$ W$^{-1}$ and $\gamma^{(A)}$=330 m$^{-1}$ W$^{-1}$ are the nonlinear coefficients. The linear coupling between two mode families is induced by $iL\kappa E^{(x)}  (x = A,P)$, where $\kappa$=84 m$^{-1}$ is linear coupling strength. $\Delta$ indicates the resonant frequency difference between the two modes, which is equal to $t_{R}\left[\left(\omega_{0}^{(A)}-\omega_{0}^{(P)}\right)-i\left(\beta_{1}^{(A)}-\beta_{1}^{(P)}\right) \frac{\partial}{\partial \tau}\right]$, where $\omega_{0}^{(A)}$ is the resonance frequency, $\beta_{1}^{(P)}$=12.280 ns/m and $\beta_{1}^{(A)}$=13.438 ns/m are the first-order dispersion coefficients. All paramaters used for simulation can be extracted from the experiment data. The $\omega_{0}^{(A)}-\omega_{0}^{(P)}$ is tuned to simulate the change of AMX. To get the results shown in Fig. \ref{fig2}b, the $t_{R}\left(\omega_{0}^{(A)}-\omega_{0}^{(P)}\right)$ is set to 0.81. In Fig. \ref{fig2}d, the $t_{R}\left(\omega_{0}^{(A)}-\omega_{0}^{(P)}\right)$ is set to -0.01, -1.40, -1.96 and -2.49 for dark-pulses with different FSR.

\vspace{3pt}
\noindent\textbf{Data availability}\\
The data that supports the plots within this paper and other findings of this study are available from the corresponding authors upon reasonable request. 

\vspace{3pt}
\noindent\textbf{Code availability}\\
The codes that support the findings of this study are available from the corresponding authors upon reasonable request.

\end{footnotesize}
\vspace{20pt}

\bibliography{refqf.bib, REF.bib}


\vspace{12pt}
\begin{footnotesize}

\vspace{6pt}
\noindent \textbf{Acknowledgment}

\noindent 
The authors thank Weiqiang Wang, Xinyu Wang, Xiaoxiao Xue for fruitful discussions, Theodore J Morin for helpful commentary on the manuscript, and Shenzhen PhotonX Technology Co., Ltd., for laser packaging support. The UCSB nano-fabrication facility was used.

\vspace{6pt}
\noindent \textbf{Author contributions}

\noindent 
The experiments were conceived by H.S., L.C. and Q.-F.Y. The devices were designed by H.S., L.C., and W.X. The microcomb simulation and modelling is conducted by B.S. The AlGaAsOI microresonators are fabricated by W.X. and L.C. The dispersion and noise are measured by C.L. and Q.-F.Y. Other characterizations are conducted by H.S. and B.S, with the assistance from R.C., X.Z., M.J., Y.T. and Z.T. The results are analyzed by H.S., C.L., and B.S. All authors participated in writing the manuscript. The project was coordinated by H.S., L.C. and Q.-F.Y. under the supervision of S.Y., X.W and J.E.B. 

\vspace{6pt}
\noindent
\textbf{Additional information} 

\noindent Supplementary information is available in the online version of the paper. Reprints and permissions information is available online. Correspondence and requests for materials should be addressed to X.W. and J.E.B.

\vspace{6pt}
\noindent \textbf{Competing financial interests} 

\noindent The authors declare no competing financial interests.
\end{footnotesize}

\end{document}